\documentstyle[12pt]{article}

\begin{document}
\vspace*{3cm}
\begin{center}
{\large{\bf CIRCULAR STRINGS IN DE SITTER SPACETIME}}
\footnote{Talk presented at the 2\`{e}me Journ\'{e}e Cosmologie in
Paris, June
2-4 1994. To appear in the \hspace*{6mm}proceedings.}\\
\vspace*{2.5cm}
ARNE L. LARSEN\\
\vspace*{5mm}
{\it NORDITA, Blegdamsvej 17, Dk-2100 Copenhagen \O, Denmark.}
\end{center}
\vspace{.4cm}
\begin{abstract}
\hspace*{-6mm}String
propagation is investigated in de Sitter and black hole backgrounds
using both exact and approximative methods. The circular string evolution
in de Sitter space is discussed in detail with respect to energy and pressure,
mathematical solution and physical interpretation, multi-string solutions etc.
We compare with the circular string evolution in the $2+1$ dimensional black
hole anti de Sitter spacetime and in the equatorial plane of ordinary
$3+1$ dimensional stationary
axially symmetric spacetime solutions of Einstein general relativity.

Using an approximative string perturbation approach we consider also generic
string evolution and propagation in all these curved spacetimes.
\end{abstract}
\newpage
\section{Introduction}
The classical and quantum propagation of strings in curved spacetimes has
attracted a great deal of interest in recent years. The main complication, as
compared to the case of flat Minkowski space, is related to the non-linearity
of the equations of motion. It makes it possible to obtain the complete
analytic solution only in a very few special cases like conical spacetime
\cite{ve1} and plane wave/shock wave backgrounds \cite{ve2}. There are however
also very general results concerning integrability and solvability for
maximally symmetric spacetimes \cite{mic1} and gauged WZW models
\cite{bar}. These are the exceptional cases, generally the string equations
of motion in curved spacetimes are not integrable and even if they are, it
is usually an extremely difficult task to actually separate the equations,
integrate them and finally write down the complete solution in closed form.
Fortunately there are different ways to "attack" a system of coupled
non-linear partial differential equations. Besides numerical methods, which
will not be considered here, there are essentially two ways to proceed:
\vskip 6pt
\hspace*{-6mm}{\bf 1. Approximative methods.}
An approximative method was first developed by de Vega and S\'{a}nchez
\cite{ve3} (see also \cite{men,fro1,all1}). The idea was to expand
the target space coordinates around a special solution, usually taken to
be the string center of mass, and then to (try to) solve the string equations
of motion and constraints order by order. This can be done, at least up to
first or second order in the expansion, for most of the known gravitational
and cosmological spacetimes. For a review of the method and its applications,
see for instance \cite{ve5}.

Another approximative method, which is however not un-related to the first one,
was considered in Refs.\cite{ven}. It consists of a large scale-factor
expansion in spatially flat FRW-universes and an important result of the
analysis was the discovery of "extremely unstable strings" of negative
pressure in inflationary universes.
\vskip 6pt
\hspace*{-6mm}{\bf 2. Exact methods.}
As already mentioned there are a few examples where the equations of motion
and constraints can actually be solved completely. If this is not possible one
can try to find special, but exact, solutions by making an ansatz. If the
ansatz is properly chosen, i.e. such that it exploits the symmetries of
the spacetime, it may reduce the original system of coupled non-linear
partial differential equations to something much simpler, for instance
decoupled non-linear ordinary differential equations, and it may then be
possible to find the complete solution of this reduced system. This method
has been used for stationary open strings \cite{fro1,fro2}
and for circular strings in a variety
of curved spacetimes \cite{all1,mic2,all2,all3,verb}. When using this
method, one
somehow has to argue that the special solutions under consideration are
representative in the sense that their physical properties are more general
than the solutions themselves. This is done by comparing with the results
obtained from the approximative methods, so it is very important to have,
and to use, both kinds of methods.
\vskip 12pt
\hspace*{-6mm}In this talk I will present material bases on both exact and
approximative methods and I will try to take a physical point of view and
address the question: what are the strings actually doing in the curved
spacetimes?

The talk is organized as follows. In Section 2 we consider circular strings
in stationary axially symmetric backgrounds. The main result is the effective
potential determining the circular string radius. The main part of the talk
concerns circular strings in de Sitter space, as indicated by the title. The
material is based on Refs.\cite{all2,all3,all4} and is presented in Section 3.
We first consider the energy and pressure of the different types of strings.
We then turn to the mathematical solutions and their physical interpretation,
following closely Ref.\cite{all2}. Finally we consider briefly the
propagation of small perturbations around the circular strings. In Section 4
we consider circular strings in the background of the $2+1$ dimensional black
hole anti de Sitter spacetime, recently found by Ba\~{n}ados et. al. \cite{ban}
and in Section 5 we compare with circular strings in ordinary black hole and
anti de Sitter spacetimes. In Section 6 we first refine the approximative
string perturbation approach of de Vega and S\'{a}nchez \cite{ve3} by
eliminating, from the beginning, the perturbations in the direction of the
geodesic of the string center of mass. We then discuss some of the results
obtained in Ref.\cite{all1}, using this method, and we compare with the results
obtained for circular strings.

The main conclusions of the talk are compactly summarized in
tables I,II,III.
\section{Circular Strings in Stationary Axially Symmetric Backgrounds}
\setcounter{equation}{0}
In this section we consider circular strings embedded in stationary axially
symmetric backgrounds. The
analysis is carried out in $2+1$ dimensions, but
the results will hold for the equatorial plane of higher dimensional
backgrounds as well. To be more specific we consider the following line
element:
\begin{equation}
ds^2=g_{tt}(r)dt^2+g_{rr}(r)dr^2+2g_{t\phi}(r)dtd\phi+g_{\phi\phi}(r)d\phi^2,
\end{equation}
that will be general enough for our purposes here.

The circular string ansatz, consistent with the symmetries of the background,
is taken to be:
\begin{equation}
t=t(\tau),\;\;\;r=r(\tau),\;\;\;\phi=\sigma+f(\tau),
\end{equation}
where the three functions $t(\tau),\;r(\tau)$ and
$f(\tau)$ are to be determined
by the equations of motion and constraints. The equations of motion
lead to:
\begin{eqnarray}
\ddot{t}+2\Gamma^t_{tr}\dot{t}\dot{r}+2\Gamma^t_{\phi r}\dot{r}
\dot{f}=0,\nonumber
\end{eqnarray}
\begin{eqnarray}
\ddot{r}+\Gamma^r_{rr}\dot{r}^2+\Gamma^r_{tt}\dot{t}^2+\Gamma^r_{\phi\phi}
(\dot{f}^2-1)+2\Gamma^r_{t\phi}\dot{t}\dot{f}=0,
\end{eqnarray}
\begin{eqnarray}
\ddot{f}+2\Gamma^\phi_{tr}\dot{t}\dot{r}+2\Gamma^\phi_{\phi r}\dot{f}
\dot{r}=0,\nonumber
\end{eqnarray}
while the constraints become:
\begin{eqnarray}
g_{tt}\dot{t}^2+g_{rr}\dot{r}^2+g_{\phi\phi}(\dot{f}^2+1)+2g_{t\phi}
\dot{t}\dot{f}=0,\nonumber
\end{eqnarray}
\begin{eqnarray}
g_{t\phi}\dot{t}+g_{\phi\phi}\dot{f}=0.
\end{eqnarray}
This system of second order ordinary differential equations and constraints is
most easily described as a Hamiltonian system:
\begin{equation}
{\cal H}=\frac{1}{2}g^{tt}P^2_t+\frac{1}{2}g^{rr}P^2_r+\frac{1}{2}
g^{\phi\phi}P^2_\phi+g^{t\phi}P_tP_\phi+\frac{1}{2}g_{\phi\phi},
\end{equation}
supplemented by the constraints:
\begin{equation}
{\cal H}=0,\;\;\;\;P_\phi=0.
\end{equation}
The function $f(\tau)$ introduced in Eq. (2.2) does not represent a physical
degree of freedom. It describes the "longitudinal" rotation of the circular
string and is therefore a pure gauge artifact. This interpretation is
consistent with Eq. (2.6) saying that there is no angular momemtum $P_\phi$.

The Hamilton equations of the two cyclic coordinates $t$ and $f$ are:
\begin{equation}
\dot{f}=g^{\phi\phi}P_\phi+g^{t\phi}P_t,\;\;\;\;
\dot{t}=g^{tt}P_t+g^{t\phi}P_\phi,
\end{equation}
as well as:
\begin{equation}
P_\phi=\mbox{const.}=0,\;\;\;\;P_t=\mbox{const.}\equiv -E,
\end{equation}
where $E$ is an integration constant and we used Eq. (2.6). The two functions
$t(\tau)$ and $f(\tau)$ are then determined by:
\begin{equation}
\dot{f}=-Eg^{t\phi},
\end{equation}
\begin{equation}
\dot{t}=-Eg^{tt},
\end{equation}
that can be integrated provided $r(\tau)$ is known. Using Eq. (2.6) and
Eqs. (2.9)-(2.10), the Hamilton equation of $r$ becomes after one integration:
\begin{equation}
\dot{r}^2+V(r)=0;\;\;\;\;\;\;\;\;V(r)=g^{rr}(E^2g^{tt}+g_{\phi\phi}),
\end{equation}
so that $r(\tau)$ can be obtained by inversion of:
\begin{equation}
\tau-\tau_o=\int_{r_o}^r\frac{dx}{\sqrt{-g^{rr}(x)[E^2g^{tt}(x)+
g_{\phi\phi}(x)]}}.
\end{equation}
For the cases that we will consider in the following, Eq. (2.12) will be solved
in terms of either elementary or elliptic functions. By definition of the
potential $V(r),$ Eq. (2.11), the dynamics takes place at the $r-$axis in a
$(r,V(r))-$diagram. The first thing to do for a stationary axially symmetric
background under consideration, is therefore to find the zeros of the
potential. Then one can try to solve the equations of motion and finally
extract the physics of the problem
\vskip 6pt
\hspace*{-8mm}
We close this section by the following interesting observation: insertion
of the ansatz, Eq. (2.2), using the results of
Eqs. (2.9)-(2.11) in the line element, Eq. (2.1),
leads to:
\begin{equation}
ds^2=g_{\phi\phi}(d\sigma^2-d\tau^2).
\end{equation}
We can then identify the invariant string size as:
\begin{equation}
S(\tau)=\sqrt{g_{\phi\phi}(r(\tau))}
\end{equation}
\section{Circular Strings in de Sitter Space}
\setcounter{equation}{0}
$2+1$ dimensional de Sitter space is a 3 dimensional hyperboloid embedded in
4 dimensional Minkowski space:
\begin{equation}
ds^2=\frac{1}{H^2}[-(dq^0)^2+\sum_{i=1}^{3}(dq^{i})^2],\;\;\;\;\;\;\;\;
\eta_{\mu\nu}q^\mu q^\nu=1.
\end{equation}
The hyperboloid coordinates cover the whole manifold contrary to
the comoving coordinates:
\begin{equation}
ds^2=-(dX^0)^2+e^{2HX^0}[dR^2+R^2d\phi^2],
\end{equation}
and the static coordinates:
\begin{equation}
ds^2=-(1-H^2r^2)dt^2+\frac{dr^2}{1-H^2r^2}+r^2 d\phi^2.
\end{equation}
For our purposes it is however most convenient to start out with the
static coordinates since then we can use directly the results of Section 2
for the circular strings. Afterwards, we then have to transform the solutions
back to the hyperboloid using the appropriate coordinate transformations inside
and outside the horizon.

In static coordinates we can now immediately write down the
potential, Eq. (2.11),
determining the dynamics of circular strings in de Sitter space:
\begin{equation}
V(r)=r^2-H^2r^4-E^2,
\end{equation}
see Fig.1. For $4H^2E^2<1$ the top of the potential is above the $r-$axis and
there will be oscillating solutions to the left of the barrier
and bouncing solutions
to the right. The bouncing solutions will re-expand towards infinity.
For $4H^2E^2>1,$ on the other hand, the top of the potential is below the
$r-$axis and therefore the potential does not act as a barrier. Strings
can expand from $r=0$ towards infinity and very large strings can collapse.
Finally if $4H^2E^2=1$ there is a (unstable) stationary string at the top of
the potential as well as expanding and collapsing strings. It is instructive to
compare with the circular string dynamics in Minkowski space. In that case
the potential is given by $V(r)=r^2-E^2$ and only oscillating strings exist.
Physically the difference is not so difficult to understand. In flat
Minkowski space the dynamics of a circular string is determined by the string
tension only, that will always try to contract the string. This leads to
strings oscillating between $r=0$ and some maximal size depending on the
energy. In de Sitter space
there is an opposite effect namely the expansion of the universe, that will
try to expand the string. This gives rise to a region where the tension is
strongest (to the left of the barrier), a region where the expansion of the
universe is strongest (to the right of the barrier) and there may even be an
exact balance of the two opposite forces, implying the existence of a
stationary solution.

Before coming to the mathematical solution of the equations of motion, we
consider the energy and pressure of the strings.
The spacetime string
energy-momentum tensor is:
\begin{equation}
\sqrt{-g}T^{\mu\nu}(X)=\frac{1}{2\pi\alpha'}\int d\sigma d\tau(\dot{X}^\mu
\dot{X}^\nu-X'^\mu X'^\nu)\delta^{(3)}(X-X(\tau,\sigma)).
\end{equation}
After integration over a spatial volume that completely encloses the string,
the
energy-momentum tensor takes the form of a fluid:
\begin{equation}
T^\mu_\nu=\mbox{diag.}(-En.,\;Pr.,\;Pr.),
\end{equation}
where, in the comoving coordinates introduced in Eq. (3.2):
\begin{equation}
En.(X)=\frac{1}{\alpha'}\dot{X}^0,
\end{equation}
\begin{equation}
Pr.(X)=\frac{1}{2\alpha'}\frac{e^{2HX^0}(\dot{R}^2-R^2)}{\dot{X}^0},
\end{equation}
represent the energy and pressure, respectively.
{}From these expressions we can actually get a lot of information about the
energy
and pressure, without using the explicit time evolution of the strings.
{}From the coordinate transformations between comoving and static coordinates
we get the explicit expressions \cite{all4}:
\begin{equation}
H\alpha'En.=\frac{H^2r\dot{r}-HE}{H^2r^2-1},
\end{equation}
\begin{equation}
H\alpha'Pr.=\frac{(H^4r^4-2H^2r^2-H^2E^2)H^2r\dot{r}+
(3H^4r^4-2H^2r^2+H^2E^2)HE}{2(1-H^2r^2)(H^4r^4+
H^2E^2)}.
\end{equation}
Both the energy and the pressure now depend on the string radius $r$ and the
velocity $\dot{r}.$ The latter can however
be eliminated using Eq. (2.11) and Eq. (3.4).

Let us first consider a string expanding from $r=0$ towards infinity.
This corresponds to a string with $\dot{r}>0$ and $H^2E^2>1/4$, see
Fig.1. For $r=0$ we find $En.=E/\alpha'$ and $Pr.=E/(2\alpha'),$
thus the equation
of state is $Pr.=En./2.$ This is like
ultra-relativistic matter. As the string expands, the energy soon starts to
increase while the pressure starts to decrease and becomes negative, see
Fig.2. For $r\rightarrow\infty$ we find $En.=r/\alpha'$ and
$Pr.=-r/(2\alpha'),$
thus not surprisingly, we have recovered the equation of state of extremely
unstable strings $Pr.=-En./2.$ \cite{ven}.
Now consider an oscillating string, i.e. a string
with $H^2E^2\leq 1/4$ in the region to the left of the potential
barrier, Fig.1. The equation of state near $r=0$ is the same as for the
expanding string, but now the string has a maximal radius:
\begin{equation}
Hr_{\mbox{max}}\equiv \nu=\sqrt{\frac{1-\sqrt{1-4H^2E^2}}{2}}.
\end{equation}
For $r=r_{\mbox{max}}$ we find:
\begin{equation}
H\alpha'En.=\frac{\nu}{\sqrt{1-\nu^2}}\equiv k,\;\;\;\;\;\;
H\alpha'Pr.=\frac{k}{2}(-1+\frac{2k^2}{1+k^2}),
\end{equation}
corresponding to a perfect fluid type equation of state:
\begin{equation}
Pr.=(\gamma-1)En.,\;\;\;\;\;\;\;\;\gamma=\frac{k^2}{1+k^2}+\frac{1}{2}.
\end{equation}
Notice that $k\in[0,\;1],$ with $k=0$ decribing a string at the bottom of the
potential while $k=1$ decribes a string oscillating between $r=0$ and the
top of the potential barrier (in this case the string actually only makes
one oscillation \cite{mic2,all2}). For
$k<<1$ the equation of state, Eq.(3.13), near the
maximal radius reduces to $Pr.=-En./2.$  In the other
limit $k\rightarrow 1$ we find, however,
$Pr.\rightarrow 0$ corresponding to cold matter. For the oscillating
strings we can also calculate the average values of energy and pressure by
integrating over a full period, see Fig.3. Since
a $\tau$-integral can be converted into
a $r$-integral, using Eq. (2.11) and Eq. (3.4), the
average values can be obtained without
using the exact $\tau$-dependence of the string radius. The
average energy becomes:
\begin{equation}
H\alpha'<En.>=\frac{2k}{T\sqrt{1+k^2}}\Pi(\frac{-k^2}{1+k^2},\;k),
\end{equation}
where $T$ is the period and $\Pi$ is the complete elliptic integral of the
third kind (using the notation of Gradshteyn \cite{gra}). The average pressure
is zero, thus in average the oscillating strings
describe cold matter. In Minkowski space, for comparison, it can easily be
shown \cite{all4} that the
energy is constant while the pressure depends on the string
radius. The average pressure is however zero, just as in de Sitter space.
\subsection{The Mathematical Solution}
(This and the following subsection follow closely Ref.\cite{all2}.)\\
We now come to the explicit mathematical solutions. In the general
case equations (2.11) and (3.4)
are solved by:
\begin{equation}
H^2r^2(\tau)=\wp(\tau-\tau_o)+1/3,
\end{equation}
where $\wp$ is the Weierstrass elliptic $\wp$-function \cite{abr}:
\begin{equation}
\dot{\wp}^2=4\wp^3-g_2\wp-g_3,
\end{equation}
with invariants:
\begin{equation}
g_2=4(\frac{1}{3}-H^2E^2),\;\;\;g_3=\frac{4}{3}(\frac{2}{9}-H^2E^2),
\end{equation}
and discriminant:
\begin{equation}
\Delta\equiv g_2^3-27g_3^2=16H^4E^4(1-4H^2E^2).
\end{equation}
It is convenient to
consider separately the 3 cases $H^2E^2=1/4\;(\Delta=0)$,
$H^2E^2<1/4\;(\Delta>0)$
and $H^2E^2>1/4\;(\Delta<0)$, see Fig.1.
\vskip 12pt
\hspace*{-6mm}${\bf H^2E^2=1/4}$: In this
case the Weierstrass function reduces to a hyperbolic
function and Eq. (3.15) becomes:
\begin{equation}
H^2r^2(\tau)=\frac{1}{2}(1+\sinh^{-2}(\frac{\tau-\tau_o}{\sqrt{2}})).
\end{equation}
Two real independent solutions are obtained by the choices $\tau_o=0$ and
$\tau_o=i\pi/\sqrt{2}$, respectively:
\begin{equation}
H^2r^2_-(\tau)=\frac{1}{2}\tanh^{-2}\frac{\tau}{\sqrt{2}},
\end{equation}
\begin{equation}
H^2r^2_+(\tau)=\frac{1}{2}\tanh^2\frac{\tau}{\sqrt{2}}.
\end{equation}
Notice that:
\begin{eqnarray}
&H^2r^2_-(-\infty)=\frac{1}{2},\;\;\;H^2r^2_-(0)=\infty,\;\;\;
H^2r^2_-(\infty)=\frac{1}{2},&
\nonumber\\
&H^2r^2_+(-\infty)=\frac{1}{2},\;\;\;H^2r^2_+(0)=0,\;\;\;
H^2r^2_+(\infty)=\frac{1}{2}.&
\end{eqnarray}
These are the 2 solutions originally found by de Vega, S\'{a}nchez and
Mikhailov \cite{mic2}, corresponding to $\alpha>0$ and $\alpha<0$ in
their notation, respectively.
The interpretation of these solutions as a function
of the world-sheet time $\tau$ is clear from Fig.1: the solution, Eq. (3.20),
expands from $H^2r_-^2=1/2$ towards infinity and then contracts until it
reaches its original size. The solution, Eq. (3.21), contracts
from $H^2r_+^2=1/2$
until it collapses. It then expands again until it reaches its
original size. The physical interpretation,
that is somewhat more involved, was described in
Ref.\cite{mic2} and
will be shortly reviewed in Subsection 3.2. There is actually also a
stationary solution for $H^2E^2=1/4$, i.e.
a string sitting on the top
of the potential, see Fig.1. This solution with constant string size
$S=1/(\sqrt{2}H)$ was discussed in
Ref.\cite{mic2} and will not be considered here.
\vskip 12pt
\hspace*{-6mm}${\bf H^2E^2<1/4}$: Here
two real independent solutions are obtained
by the choices $\tau_o=0$ and $\tau_o=\omega'$, respectively:
\begin{equation}
H^2r^2_-(\tau)=\wp(\tau)+1/3,
\end{equation}
\begin{equation}
H^2r^2_+(\tau)=\wp(\tau+\omega')+1/3,
\end{equation}
where $\omega'$ is the imaginary semi-period of the Weierstrass function. It
is explicitly given by \cite{abr}:
\begin{equation}
\omega'=i\frac{\sqrt{2}K'(k)}{\sqrt{1+\sqrt{1-4H^2E^2}}};\;\;\;k=\sqrt{\frac
{1-\sqrt{1-4H^2E^2}}{1+\sqrt{1-4H^2E^2}}}.
\end{equation}
Notice that:
\begin{eqnarray}
\hspace*{-6mm}H^2r^2_-(0)\hspace*{-2mm}&=&\hspace*{-2mm}\infty,
\;\;\;H^2r^2_-(\omega)=(1+\sqrt{1-4H^2E^2})/2,\;\;\;
H^2r^2_-(2\omega)=
\infty,...\nonumber\\
\hspace*{-6mm}H^2r^2_+(0)\hspace*{-2mm}&=&\hspace*{-2mm}0,\;\;\;H^2r^2_+
(\omega)=(1-\sqrt{1-4H^2E^2})/2,\;\;\;
H^2r^2_+(2\omega)=0,
...
\end{eqnarray}
where $\omega$ is the real semi-period of the Weierstrass function \cite{abr}:
\begin{equation}
\omega=\frac{\sqrt{2}K(k)}{\sqrt{1+\sqrt{1-4H^2E^2}}},
\end{equation}
and $K'$ and $K$ are the complete elliptic integrals of first kind.
The interpretation of these solutions as a function of $\tau$
is clear from Fig.1: the solution, Eq. (3.23), oscillates
between infinity and its
minimal size $H^2r_-^2=(1+\sqrt{1-4H^2E^2})/2$ at
the boundary of the potential, while
the solution, Eq. (3.24), oscillates between $0$ and its maximal size
$H^2r_+^2=(1-\sqrt{1-4H^2E^2})/2$. The
physical interpretation will be considered in Subsection 3.2.
\vskip 12pt
\hspace*{-6mm}${\bf H^2E^2>1/4}$: In
this last case two real independent solutions
are obtained
by the choices $\tau_o=0$ and $\tau_o=\omega_2'$, respectively:
\begin{equation}
H^2r^2_-(\tau)=\wp(\tau)+1/3,
\end{equation}
\begin{equation}
H^2r^2_+(\tau)=\wp(\tau+\omega_2')+1/3,
\end{equation}
where $\omega_2'$ takes the explicit form:
\begin{equation}
\omega_2'=i\frac{K'(\hat{k})}{\sqrt{HE}};\;\;\;\hat{k}
=\sqrt{\frac{1}{2}+\frac{1}{4HE}}.
\end{equation}
Notice that:
\begin{eqnarray}
&H^2r^2_-(0)=\infty,\;\;\;H^2r^2_-(\omega_2)=0,\;\;\;H^2r^2_-(2\omega_2)
=\infty,...&\nonumber\\
&H^2r^2_+(0)=0,\;\;\;H^2r^2_+(\omega_2)=\infty,\;\;\;H^2r^2_+(2\omega_2)=0,...&
\end{eqnarray}
where:
\begin{equation}
\omega_2=\frac{K(\hat{k})}{\sqrt{HE}}.
\end{equation}
It should be stressed that in this case the
primitive semi-periods are $\hat{\omega}=(\omega_2-\omega'_2)/2$ and
$\hat{\omega}'=(\omega_2+\omega'_2)/2$, i.e.
$(2\hat{\omega},2\hat{\omega}')$ spans a
fundamental period
parallelogram in the complex plane.

The interpretation of the solutions, Eqs. (3.28)-(3.29),
as a function of $\tau$
follows from Fig.1: both of them oscillates between zero size
(collapse) and infinite size (instability). The
physical interpretations will follow in Subsection 3.2.
\subsection{The Physical Interpretation}
${\bf H^2E^2=1/4}$: We first consider
the $r_-$-solution, Eq. (3.20). The hyperboloid time is obtained by integrating
Eq. (2.10) and transforming back to the hyperboloid coordinates:
\begin{equation}
q^0_-(\tau)=\sinh\tau-\frac{1}{\sqrt{2}}\cosh\tau\coth\frac{\tau}{\sqrt{2}}.
\end{equation}
When we plot this function (Fig.4.) we see that the string solution actually
describes 2 strings (I and II) \cite{mic2}, since
$\tau$ is a two-valued function of $q^0_-$.
For both strings the invariant string size is:
\begin{equation}
S_-=\frac{1}{\sqrt{2}H}\coth\mid\frac{\tau}{\sqrt{2}}\mid,
\end{equation}
but string I corresponds to $\tau\in\;]-\infty,0[\;$ and string II to
$\tau\in\;]0,\infty[\;$. Therefore, $q^0_-\rightarrow\infty$ corresponds to
$\tau\rightarrow 0_-$ for string I, but to $\tau\rightarrow\infty$ for string
II. More generally, when $q^0_-\rightarrow\infty$ the invariant size
grows indefinitely for string I, while it approaches a constant
value for string II. We conclude that string I is an unstable string
for $q^0_-\rightarrow\infty$, while string II is a stable string. More details
about the connection between hyperboloid time and world-sheet time for these
solutions can be found in Ref.\cite{mic2}.
Let us consider now the comoving time
of the solution $r_-$ in a little more detail:
\begin{equation}
X^0_-(\tau)=\frac{1}{H}(\tau+\log\mid\frac{1}{\sqrt{2}}\coth\frac{\tau}{\sqrt{2}}
-1\mid).
\end{equation}
When we plot this function (Fig.4.) we find that
$\tau$ is a three-valued function of $X^0_-$. What happens is that
the time interval $\tau\in\;]0,\infty[\;$ for
string II splits into two parts. These features
are easily understood when returning to the effective potential, Fig.1.:
String I starts at $H^2r_-^2=1/2$ for $\tau=HX^0_-=-\infty$, it
then expands through
the horizon $H^2r_-^2=1$ at:
\begin{equation}
\tau=-\sqrt{2}\log(1+\sqrt{2}),\;\;\;HX^0_-=\log 2-\sqrt{2}\log(1+\sqrt{2})
\end{equation}
and continues towards infinity for $\tau\rightarrow 0_-,\;HX^0_-
\rightarrow\infty$.
String II starts at infinity for $\tau=0_+,\;HX^0_-=\infty$ and
contracts through
the horizon at:
\begin{equation}
\tau=\sqrt{2}\log(1+\sqrt{2}),\;\;\;HX^0_-=-\infty.
\end{equation}
This behaviour, approaching the horizon from the outside,
corresponds to the going backwards in comoving time-part of Fig.4. String II
then continues contracting from $H^2r_-^2=1$ at:
\begin{equation}
\tau=\sqrt{2}\log(1+\sqrt{2}),\;\;\;HX^0_-=-\infty,
\end{equation}
until it reaches $H^2r^2_-=1/2$ at $\tau=HX^0_-=\infty$.
We now consider briefly the $r_+$-solution, Eq. (3.21). In this case
the hyperboloid time is given by:
\begin{equation}
q^0_+(\tau)=\sinh\tau-\frac{1}{\sqrt{2}}\cosh\tau\tanh\frac{\tau}{\sqrt{2}},
\end{equation}
which
is a monotonically increasing function of $\tau$. The $r_+$-solution therefore
describes only one string. The proper size is
given by:
\begin{equation}
S_+=\frac{1}{\sqrt{2}H}\tanh\mid\frac{\tau}{\sqrt{2}}\mid
\end{equation}
We can also express this solution in terms of the comoving time:
\begin{equation}
X^0_+(\tau)=\frac{1}{H}
(\tau+\log(1-\frac{1}{\sqrt{2}}\tanh\frac{\tau}{\sqrt{2}})),
\end{equation}
but since everything now takes place well inside the horizon, this will not
really give us more insight. The string starts with $H^2r_+^2=1/2$ for $\tau=
HX^0_+=-\infty.$ It then contracts until
it collapses for $\tau=HX^0_+=0$ and expands
again and eventually reaches $H^2r_+^2=1/2$ for
$\tau=HX^0_+=\infty.$
\vskip 12pt
\hspace*{-6mm}${\bf H^2E^2<1/4}$:
Using the notation
introduced in Eqs. (3.11)-(3.12)
the solutions, Eqs. (3.23)-(3.24), can be written as:
\begin{equation}
H^2r^2_-(\tau)=\frac{\mu^2}{{\mbox{sn}}^2[\mu\tau\mid k]},
\end{equation}
\begin{equation}
H^2r^2_+(\tau)=\nu^2 {\mbox{sn}}^2[\mu\tau\mid k],
\end{equation}
where $\mu=\sqrt{1-\nu^2}.$ Consider first the $r_-$-solution, Eq. (3.42). It
is clear from Eq. (3.26) and the
periodicity in general that we have infinitely many branches $[0,2\omega],\;
[2\omega,4\omega],\;...$. We will see in a moment that each of these
branches actually corresponds to one string, that is, the $r_-$-solution
describes
infinitely many strings. For that purpose we will need the
hyperboloid time and the
comoving time as a function of $\tau$. Both of them are expressed in terms of
the static coordinate time $t$, that is obtained by integrating Eq. (2.10):
\begin{equation}
Ht_-(\tau)=\zeta(x/\mu)\tau+
\frac{1}{2}\log\mid\frac{\sigma(\tau-x/\mu)}{\sigma(\tau+x/\mu)}\mid,
\end{equation}
where $\zeta$ and $\sigma$ are the Weierstrass $\zeta$
and $\sigma$-functions \cite{abr}
and $x$ is a real constant obeying $\mbox{sn}[x\mid k]=\mu$, i.e. $x$ is
expressed
as an incomplete elliptic integral of the first kind. The
expression, Eq.(3.44), can be further rewritten in terms of
theta-functions \cite{abr}:
\begin{equation}
Ht_-(\tau)=\frac{\mu\tau\pi}{2K}\frac{\vartheta'_1}
{\vartheta_1}(\frac{\pi x}{2K})+
\frac{1}{2}\log\mid\frac{\vartheta_1(\frac{\pi(\mu\tau-x)}{2K})}
{\vartheta_1(\frac{\pi(\mu\tau+x)}{2K})}\mid,
\end{equation}
and finally as:
\begin{equation}
Ht_-(\tau)=\frac{1}{2}\log\mid\frac{\sin(\frac{\pi(\mu\tau-x)}
{2K})}{\sin(\frac{\pi(\mu\tau+x)}{2K})}\mid+\frac{\mu\tau\pi}{2K}
\frac{\vartheta'_1}{\vartheta_1}
(\frac{\pi x}{2K})
-2\sum_{n=1}^\infty\frac{q^{2n}}{n(1-q^{2n})}
\sin(\frac{n\pi\mu\tau}{K})\sin(\frac{n\pi x}{K}),
\end{equation}
where $q=e^{-\pi K'/K}$.
In the latter expression we have isolated all the real
singularities in the first term. To be more specific we see that the static
coordinate time is singular for $\mu\tau\rightarrow 2KN\pm x$, where
$N$ is an integer, with the
asymptotic behaviour:
\begin{equation}
t_-(\tau)\rightarrow\pm\frac{1}{2H}\log\mid\mu\tau -2KN\mp x\mid;\;\;\;\;
\tau\rightarrow \frac{2K}{\mu}N\pm\frac{x}{\mu}.
\end{equation}
On the other hand $t_-(\tau)$ is completely regular at the boundaries of
the branches, i.e. for $\tau=0,\;\pm2\omega,\;\pm4\omega,... $. These results
can be easily translated to the hyperboloid time \cite{all2}:
\begin{equation}
q^0_-(\tau)=-\frac{\Omega\vartheta'_1(0)}{2\pi}\frac{e^{\Omega\tau\frac
{\vartheta'_1}{\vartheta_1}(\Omega y)}\vartheta_1(\Omega(y-\tau))+
e^{-\Omega\tau\frac
{\vartheta'_1}{\vartheta_1}(\Omega y)}\vartheta_1(\Omega(y+\tau))}
{\vartheta_1(\Omega\tau)\vartheta_1(\Omega y)},
\end{equation}
where:
\begin{equation}
\Omega\equiv\frac{\pi\mu}{2K},\;\;\;\;y\equiv\frac{x}{\mu}.
\end{equation}
Notice that the singularities, Eq. (3.47), that originated from the zeros of
$\vartheta_1(\Omega(y\pm\tau)),$ have canceled in $q^0_-(\tau)$ so that
$q^0_-(2\omega N\pm y)$ is finite.
$q^0_-(\tau)$ blows up for $\tau=0,\;\pm2\omega,\;
\pm4\omega,...$, like:
\begin{equation}
\mid q^0_-\mid\propto\mid\frac{1}{2N\omega-\tau}\mid,
\end{equation}
where $N$ is again an integer.
This demonstrates that the world-sheet time $\tau$
is actually an infinite valued
function of $q^0_-$, and that the solution $r_-$ therefore describes
{\it infinitely
many strings} (see Fig.5). This should be compared with the
$H^2E^2=1/4$ case where
we found a solution describing two strings. In that case
the two strings were of completely different type and had completely different
physical interpretations. In the present case we find infinitely many strings
but they are all of the same type. In the branch $\tau\in[0,2\omega]$ (say)
the string starts with infinite string size at $\tau=0,\;\;q^0_-=-\infty$. It
then contracts to its minimal size $H^2r_-^2=(1+\sqrt{1-4H^2E^2})/2$
and reexpands
towards infinity at $\tau=2\omega,\;\;q^0_-=\infty$. This solution, and the
infinitely many others of the same type, are unstable strings.

The comoving time of the $r_-$-solution is given by \cite{all2}:
\begin{eqnarray}
&HX^0_-(\tau)=\Omega\tau\frac{\vartheta'_1}{\vartheta_1}(\Omega y)+
\log\mid\frac{\Omega\vartheta_1(\Omega(\tau-y))\vartheta'_1(0)}
{\pi\vartheta_1(\Omega\tau)\vartheta_1(\Omega y)}\mid &\nonumber
\end{eqnarray}
\begin{eqnarray}
&=\log\mid\frac{\Omega\sin(\Omega
(\tau-y))}{\sin(\Omega\tau)}\mid-\log\mid\frac{\pi\vartheta_1(\Omega y)}
{\vartheta'_1(0)}\mid+\Omega\tau
\frac{\vartheta'_1}{\vartheta_1}(\Omega y)&\nonumber
\end{eqnarray}
\begin{eqnarray}
&-4\sum_{m=1}^\infty\frac{q^{2m}}
{m(1-q^{2m})}
\sin(m\Omega y)\sin(m\Omega(2\tau-y)).&
\end{eqnarray}
It can be shown that $0\leq\Omega\leq 1$ and $1.246..\leq y\leq\pi/2$. It
follows that:
\begin{equation}
0<y<\omega<2\omega-y<2\omega.
\end{equation}
The comoving time, Eq. (3.51), is singular
at $\tau=0,\;y,\;2\omega$ but regular at
$\tau=2\omega-y$ and similarly in the other branches, Fig.5.
Therefore, the interpretation
of the string solution in the branch $\tau\in[0,2\omega]$ (say), as seen in
comoving coordinates, is as follows: The string starts with infinite
size at $\tau=0,\;\;HX^0_-=\infty$. It then contracts and passes the
horizon from the outside at $\tau=y,\;\;HX^0_-=-\infty$. The string now
continues contracting from the inside of the horizon at $\tau=y,\;\;
HX^0_-=-\infty$ until it reaches the minimal size at:
\begin{equation}
\tau=\omega,\;\;\;\;HX^0_-
=\frac{\pi}{2}\frac{\vartheta'_1}{\vartheta_1}(\Omega y)
+\frac{1}{2}\log\frac{1-\sqrt{1-4H^2E^2}}{2}.
\end{equation}
{}From now
on the string expands again. It passes the horizon from the inside after
finite comoving
time and continues towards infinity for $HX^0_-\rightarrow\infty$.

It is an interesting observation that the comoving time is not periodic in
$\tau$, i.e. $X^0_-(\tau)\neq X^0_-(\tau+2\omega)$, although the string size
is.
Explicitly we find:
\begin{equation}
X^0_-(\tau+2\omega)-X^0_-(\tau)=\frac{\pi}{H}
\frac{\vartheta'_1}{\vartheta_1}(\Omega y).
\end{equation}
This means that $r_-(\tau)$ really describes infinitely many strings
with different invariant size
at a given comoving time. To be more specific let us
consider a fixed comoving time $X^0_-$ and the corresponding world-sheet times:
\begin{equation}
X^0_-\equiv X^0_-(\tau_1)=X^0_-(\tau_2)=...,
\end{equation}
where $\tau_1\in[0,2\omega[,\;\tau_2\in[2\omega,4\omega[...$. Taking for
simplicity a comoving time $HX^0_->>1$ we have (see Fig.5.):
\begin{equation}
\tau_n=\frac{n\pi}{\Omega}+\epsilon_n,\;\;\;\;\epsilon_n<<1.
\end{equation}
To the lowest orders we find from Eq. (3.51):
\begin{equation}
HX^0_-=-\log\epsilon_n+n\pi\frac{\vartheta'_1}{\vartheta_1}(\Omega y)+
{\cal O}(\epsilon_n),
\end{equation}
so that:
\begin{equation}
\epsilon_n=\exp[-HX^0_-+n\pi\frac{\vartheta'_1}{\vartheta_1}(\Omega y)+...]
\end{equation}
The invariant string sizes are then:
\begin{equation}
HS_-\approx\frac{1}{\epsilon_n}=\exp[HX^0_--n\pi
\frac{\vartheta'_1}{\vartheta_1}(\Omega y)+...]
\end{equation}
i.e. they are separated by a multiplicative factor. This expression, of
course, is only valid as long as $\epsilon_n<<1$, so $n$ should not be
too large.

We now consider
the $r_+$-solution, Eq. (3.43). In this case the dynamics takes
place well inside the horizon. The possible singularities of the
hyperboloid time $q^0_+$ and the comoving time $X^0_+$ therefore coincide with
the singularities of the static coordinate time $t_+$. The static coordinate
time is again obtained from Eq. (2.10), which we first rewrite as:
\begin{equation}
H\dot{t}_+(\tau)=\frac{HE}{2/3-\wp(\tau+\omega')}=
HE\;(1+\frac{H^2E^2}{\wp(\tau)-(H^2E^2-1/3)}).
\end{equation}
Integration leads to:
\begin{equation}
Ht_+(\tau)=\tau(HE+\zeta(a))+\frac{1}{2}\log\mid\frac{\sigma(\tau-a)}
{\sigma(\tau+a)}\mid,
\end{equation}
where $a$ is a complex constant obeying $\wp(a)=H^2E^2-1/3,\;\;$ i.e.
$\mbox{sn}[a\mu\mid k]=1/\nu$. It
follows that $a\mu=iK+x$ where $x$ is real and
$\mbox{sn}[x\mid k]=\mu$. Again we can express the static coordinate time in
terms of theta-functions \cite{abr}:
\begin{equation}
Ht_+(\tau)=\tau(HE+\frac{\mu\pi}{2K}\frac{\vartheta'_4}{\vartheta_4}
(\frac{\pi x}{2K}))+\frac{1}{2}
\log\mid\frac{\vartheta_4(\frac{\pi(\mu\tau-x)}{2K})}
{\vartheta_4(\frac{\pi(\mu\tau+x)}{2K})}\mid,
\end{equation}
or in terms of the Jacobi zeta-function $zn$ \cite{gra}:
\begin{equation}
Ht_+(\tau)=\tau(HE+\mu zn(x,k))-2\sum_{n=1}^\infty
\frac{q^n}{n(1-q^{2n})}\sin(\frac{n\pi\mu\tau}{K})\sin(\frac{n\pi x}{K}),
\end{equation}
where $q=e^{-\pi K'/K}$. In this form we see explicitly that $t_+$ consists
of a linear term plus oscillating terms.
The comoving time takes the form:
\begin{eqnarray}
&HX^0_+(\tau)=\frac{1}{2}\log(1-\nu^2{\mbox{sn}}^2
[\mu\tau\mid k])+Ht_+(\tau)&\nonumber
\end{eqnarray}
\begin{eqnarray}
&=\tau[HE+\Omega\frac{\vartheta'_4}{\vartheta_4}(\Omega y)]+
\log\mid\frac{\Omega\vartheta_4(\Omega(\tau-y))\vartheta'_1(0)}
{\pi\vartheta_1(\Omega y)\vartheta_4(\Omega\tau)}\mid. &
\end{eqnarray}
Notice that the argument of the $\log$ has no real zeros:
\begin{eqnarray}
&HX^0_+(\tau)=\tau[HE+\Omega\frac{\vartheta'_4}{\vartheta_4}(\Omega y)]+
\log\mid\frac{\Omega\vartheta'_1(0)}{\pi\vartheta_1(\Omega y)}\mid &\nonumber
\end{eqnarray}
\begin{eqnarray}
&-4\sum_{n=1}^{\infty}\frac{q^n}{n(1-q^{2n})}\sin(n\pi\mu\Omega)
\sin(n\pi\Omega(2\tau-y)).&
\end{eqnarray}
The static coordinate
time and the cosmic time are therefore completely regular functions
of $\tau$, and it follows that
the string solution $r_+$, which is {\it oscillating regularly}
as a function of
world-sheet
time $\tau$, is also oscillating regularly when expressed in terms of
hyperboloid time or comoving time. This solution represents one {\it stable}
string.
\vskip 12pt
\hspace*{-6mm}${\bf H^2E^2>1/4}$: The analysis here is very similar to the
analysis of the $r_--$solution in the $H^2E^2<1/4-$case
so we shall not go into it here. The results
are summarized in Table I, together with the results from the other cases.
\newpage
\begin{centerline}
{\bf Table I}
\end{centerline}
\vskip 30pt
\hspace*{-6mm}Circular string evolution in de Sitter spacetime. For each
$H^2E^2,$ there exists two independent solutions $r_-$ and $r_+$:
\vskip 30pt
\begin{tabular}{|l|l|l|}\hline
$ $&  $ $ & $ $ \\
$\hspace{6mm}H^2E^2$ & $\hspace{27mm}r_-$ & $\hspace{28mm}r_+$\\
$ $ & $ $ & $ $ \\ \hline
$ $ & $ $ & $ $ \\
$ $ & Infinitely many different strings. & One stable oscillating string  \\
$ $ & All are unstable ($Hr_{-}^{max}=\infty$), & ($0\leq r_+\leq
r_{+}^{max}\;$; $Hr_{+}^{max}=$ \\
$<1/4$ &  and never collapse. $(Hr_{-}^{min}=$ &
$\sqrt{(1-\sqrt{1-4H^2E^2})/2}\;).$ \\
$ $ & $\sqrt{(1+\sqrt{1-4H^2E^2})/2}\;>0$). & $ $ \\
$ $ & $ $ & $ $ \\ \hline
$ $ & $ $ & $ $ \\
$ $ & Infinitely many strings. All of & Infinitely many strings similar \\
$>1/4$ & them are unstable $(Hr_{-}^{max}=\infty)$ & to $r_-$. In
this case $r_+$ is just a \\
$ $ & and they collapse to a point & time-translation of $r_-$: \\
$ $ & ($r_{-}^{min}=0$). & $r_+(\tau)=r_-(\tau+\omega_2).$ \\
$ $ & $ $ & $ $ \\ \hline
 $ $ & $ $ &  $ $ \\
$ $ & Two different and non-oscillating & One stable and non-oscillating \\
$=1/4$ & strings $r_-^{(I)}$ and $r_-^{(II)}$. $r_-^{(I)}$ is &
string (it makes only one \\
$ $ & unstable and $r_-^{(II)}$ is stable for &
oscillation). $Hr_{+}^{min}=0,$ \\
$ $ & large de Sitter radius. & $Hr_{+}^{max}=1/\sqrt{2}.$ \\
 $ $ & $ $ &  $ $ \\ \hline
\end{tabular}
\subsection{Small Perturbations Around Circular Strings}
Using the covariant approach of Frolov and Larsen \cite{fro1},
we have considered
small perturbations propagating along the circular strings in black hole and
cosmological spacetime backgrounds \cite{all3,all5}. In
the case of the $3+1$ dimensional
de Sitter space introduce two normal vectors $n^\mu_\perp$ and
$n^\mu_\parallel$ perpendicular to the string world-sheet:
\begin{equation}
n^\mu_\perp=(0,\;0,\;\frac{1}{r},\;0),\;\;\;\;\;\;\;\;n^\mu_\parallel=
(\frac{\dot{r}}{r(1-H^2r^2)},\;\frac{E}{r},\;0,\;0),
\end{equation}
and then consider only physical perturbations:
\begin{equation}
\delta x^\mu=\delta x^\perp n^\mu_\perp+\delta x^\parallel n^\mu_\parallel,
\end{equation}
where $\delta x^\perp$ and $\delta x^\parallel$ are the perturbations
as seen by an observer travelling with the circular string. After Fourier
expanding $\delta x^R$:
\begin{equation}
\delta x^R(\tau,\sigma)=\sum_n C^R_n(\tau) e^{-in\sigma},\;\;\;\;\;
R=\perp,\;\parallel
\end{equation}
it can be shown that \cite{all3}:
\begin{equation}
\ddot{C}_{n\perp}+(n^2-2H^2r^2)C_{n\perp}=0,
\end{equation}
\begin{equation}
\ddot{C}_{n\parallel}+(n^2-2H^2r^2-\frac{2E^2}{r^2})C_{n\parallel}=0,
\end{equation}
determining the comoving perturbations. These equations have been discussed in
detail in Refs.\cite{all3,all5}, so we shall just give one simple result here.
For $r\rightarrow\infty$ the brackets in Eqs. (3.69)-(3.70) become
negative. This
means that the perturbations develop imaginary frequencies and grow
indefinitely. However, by considering the detailed solutions it turns out
that the perturbations grow with the same rate as the radius of the
underlying circular string (which by the way grows with the same rate as the
universe) so although the perturbations grow, the circular shape of the
string is actually stable. More details can be found in Refs.\cite{all3,all5}.
\section{Circular Strings in the 2+1 BH-ADS Spacetime}
\setcounter{equation}{0}
We now consider the circular string dynamics in the 2+1 black hole anti
de Sitter (BH-ADS) spacetime recently found by Ba\~{n}ados et. al. \cite{ban}.
This spacetime background has arised much interest recently. It
describes a two-parameter family (mass $M$ and angular momentum $J$) of black
holes in 2+1 dimensional general relativity with metric:
\begin{equation}
ds^2=(M-\frac{r^2}{l^2})dt^2+(\frac{r^2}{l^2}-M+\frac{J^2}{4r^2})^{-1}dr^2
-Jdtd\phi+r^2 d\phi^2.
\end{equation}
It has two horizons
$r_\pm=\sqrt{\frac{Ml^2}{2}\pm\frac{l}{2}\sqrt{M^2 l^2-J^2}}$ and a
static limit $r_{\mbox{erg}}=\sqrt{M}l,$ defining
an ergosphere, as for ordinary Kerr
black holes.
Using the general formalism of Section 2, we can immediately read off the
potential (see Fig.6.):
\begin{equation}
V(r)=r^2(\frac{r^2}{l^2}-M)+\frac{J^2}{4}-E^2.
\end{equation}
The potential, Eq. (4.2), has a
global minimum between the two horizons:
\begin{equation}
V_{\mbox{min}}=V(\sqrt{\frac{Ml^2}{2}}\;)=-\frac{1}{4}(M^2l^2-J^2+4E^2)<0,
\end{equation}
which is always negative, since we only consider the case when $Ml^2\geq J^2$
(otherwise there are no horizons). For large values of $r$ the
potential goes as $r^4$ and at $r=0$ we have:
\begin{equation}
V(0)=\frac{J^2}{4}-E^2,
\end{equation}
that can be either positive, negative or zero. Notice also that the potential
vanishes provided:
\begin{equation}
V(r_0)=0\;\;\Leftrightarrow\;\;r_{01,2}=\sqrt{\frac{Ml^2}{2}\pm
\frac{l}{2}\sqrt{M^2l^2-J^2+4E^2}}.
\end{equation}
There are therefore three fundamentally different types of solutions.
\vskip 6pt
\hspace*{-6mm}{\bf (i)}: For $J^2>4E^2$ there are two positive-$r$ zeros
of the
potential (Fig.6a). The smallest zero is located between the inner horizon
and $r=0$, while the other zero is between the outer horizon and the static
limit. Therefore, this string solution never comes outside the static limit. On
the other hand it never falls into $r=0$. The mathematical solution
oscillating between these two positive zeros of the potential may be
interpreted as a string travelling between the different universes described by
the maximal analytic extension of the spacetime (the Penrose diagram of
the $2+1$ dimensional BH-ADS spacetime is discussed in Refs.\cite{ban2,wel}).
Such type of
circular string solutions also exist in other stringy
black hole backgrounds \cite{ini}.
\vskip 6pt
\hspace*{-6mm}{\bf (ii)}: For $J^2<4E^2$ there is only one positive-$r$
zero of the potential, which
is always located outside the static limit (Fig.6b). The
potential is negative for $r=0$, so there is no barrier preventing the string
from collapsing into $r=0$. By suitably fixing the initial conditions the
string starts with its maximal size outside the static limit at $\tau=0$. It
then contracts through the ergosphere and the two horizons and eventually
falls into $r=0$. If $J\neq 0$ it may however still be possible to continue
this solution into another universe as in the case {\bf (i)}.
\vskip 6pt
\hspace*{-6mm}{\bf (iii)}: $J^2=4E^2$ is the limiting case where the maximal
string radius equals the static limit. The potential is exactly zero for
$r=0$ so also in this case the string contracts through the two horizons and
eventually falls into $r=0$.
\vskip 6pt
\hspace*{-6mm}The exact
and complete mathematical solution can be obtained in terms of
elementary or elliptic functions, the details
can be found in Ref.\cite{all1}. In
all cases we find only bounded string size solutions and no multi-string
solutions. See also Table II.
\section{More Spacetimes}
\setcounter{equation}{0}
We will now compare the circular strings in the
$2+1$ dimensional BH-ADS spacetime and in the equatorial plane
of ordinary $3+1$
dimensional black holes. In the most general case it is natural
to compare the spacetime
metric, Eq. (4.1), with the ordinary $3+1$ dimensional Kerr anti de
Sitter spacetime with metric components:
\begin{eqnarray}
g_{tt}=\frac{a^2\Delta_\theta\sin^2\theta-\Delta_r}{\rho^2},\;\;\;\;g_{rr}=
\frac{\rho^2}{\Delta_r},\;\;\;\;g_{t\phi}=(\Delta_r-(r^2+a^2)\Delta_\theta)
\frac{a\sin^2\theta}{\Delta_o\rho^2},\nonumber
\end{eqnarray}
\begin{eqnarray}
g_{\phi\phi}=\left( \Delta_\theta(r^2+a^2)^2-a^2\Delta_r\sin^2\theta\right)
\frac{\sin^2\theta}{\Delta^2_o\rho^2},\;\;\;\;g_{\theta\theta}=\frac{\rho^2}
{\Delta_\theta},
\end{eqnarray}
where we have introduced the notation:
\begin{eqnarray}
&\Delta_r=(1-\frac{1}{3}\Lambda r^2)(r^2+a^2)-2Mr,\;\;\;\;\Delta_\theta=
1+\frac{1}{3}\Lambda a^2\cos^2\theta,&\nonumber\\
&\Delta_o=1+\frac{1}{3}\Lambda a^2,\;\;\;\;\rho^2=r^2+a^2\cos^2\theta.&
\end{eqnarray}
Here the mass is represented by $M$ while $a$ is the specific angular momentum,
and a positive $\Lambda$ corresponds to de Sitter while a negative $\Lambda$
corresponds to anti de Sitter spacetime. In
the equatorial plane $(\theta=\pi/2)$ the
metric, Eq. (5.1), is in the general form of Eq. (2.1) so that we can use the
analysis of Section 2. In the
most general case the potential is given by:
\begin{eqnarray}
V(r)\hspace*{-2mm}&=&\hspace*{-2mm}-\frac
{\Lambda}{3\Delta_o}r^4+\frac{1-2\Lambda a^2/3}{\Delta_o}r^2-
\frac{2M(1+2\Lambda a^2/3)}{\Delta^2_o}r\nonumber\\
\hspace*{-2mm}&+&\hspace*{-2mm}\frac{2a^2-\Lambda a^2/3-E^2
\Delta^2_o}{\Delta_o}
-\frac{4M\Lambda a^4}{3\Delta^2_o}\frac{1}{r}\nonumber\\
\hspace*{-2mm}&+&\hspace*{-2mm}\frac{a^2(\Delta_o(a^2-E^2\Delta^2_o)
-4M^2)}{\Delta^2_o}\frac{1}{r^2}+
\frac{2Ma^2(a^2-E^2\Delta^2_o)}{\Delta^2_o}\frac{1}{r^3}
\end{eqnarray}
i.e. the potential covers seven powers in $r$. The general
solution will therefore
involve higher genus elliptic functions.
It is furthermore very complicated to deduce the physical properties of
the circular strings from the shape of the potential (the zeros etc.)
since the invariant string size defined in
Eq. (2.14) is non-trivially connected to $r:$
\begin{equation}
S(\tau)=\sqrt{\frac{r^2+a^2}{1+\Lambda a^2/3}+\frac{2M}{r}
\frac{a^2}{(1+\Lambda a^2/3)^2}}.
\end{equation}
We have exactly solved the string dynamics in a number of spacetimes of the
form, Eq. (5.1) \cite{all1}. In the
cases of Minkowski space, anti de Sitter space,
Schwarzschild and Schwarzschild anti de Sitter space, Fig.7., there are
only bounded string size solutions and no multi-string solutions. In
Schwarzschild de Sitter space, on the other hand, the dynamics outside the
Schwarzschild horizon is similar to the dynamics in "pure" de Sitter space, so
we find the complicated spectrum of oscillating, expanding and contracting
strings and the multi-string solutions, see Table II.
\section{Perturbations Around the String Center of Mass}
\setcounter{equation}{0}
To obtain more insight about the string propagation in all these curved
spacetimes we solved the string equations of motion and constraints
by considering perturbations around the exact
string center of mass solution, refining the approach originally
developed by de Vega and S\'{a}nchez \cite{ve3}. In
an arbitrary curved spacetime of dimension $D$, the string equations of
motion and constraints, in the conformal gauge, take the form:
\begin{equation}
\ddot{x}^\mu-x''^\mu+\Gamma^\mu_{\rho\sigma}(\dot{x}^\rho\dot{x}^\sigma-
x'^\rho x'^\sigma)=0,
\end{equation}
\begin{equation}
g_{\mu\nu}\dot{x}^\mu x'^\nu=g_{\mu\nu}(\dot{x}^\mu\dot{x}^\nu+x'^\mu
x'^\nu)=0,
\end{equation}
for $\mu=0,1,...,(D-1)$ and prime and dot represent derivative with
respect to $\sigma$ and $\tau,$ respectively. Consider
first the equations of motion, Eq. (6.1). A particular solution is
provided by the string center of mass $q^\mu(\tau)$:
\begin{equation}
\ddot{q}^\mu+\Gamma^\mu_{\rho\sigma}\dot{q}^\rho\dot{q}^\sigma=0.
\end{equation}
Then a perturbative series around this solution is developed:
\begin{equation}
x^\mu(\tau,\sigma)=q^\mu(\tau)+\eta^\mu(\tau,\sigma)+\xi^\mu(\tau,\sigma)+...
\end{equation}
After insertion of Eq. (6.4) in Eq. (6.1), the equations of motion are to
be solved order by order in the expansion.

To zeroth order we just get Eq. (6.3). To first order we find:
\begin{equation}
\ddot{\eta}^\mu+\Gamma^\mu_{\rho\sigma,\lambda}\dot{q}^\rho\dot{q}^\sigma
\eta^\lambda+2\Gamma^\mu_{\rho\sigma}\dot{q}^\rho\dot{\eta}^\sigma-
\eta''^\mu=0.
\end{equation}
The first three terms can be written in covariant form \cite{men}, c.f. the
ordinary geodesic deviation equation:
\begin{equation}
\dot{q}^\lambda\nabla_\lambda(\dot{q}^\delta\nabla_\delta\eta^\mu)-
R^\mu_{\epsilon\delta\lambda}\dot{q}^\epsilon\dot{q}^\delta\eta^\lambda-
\eta''^\mu=0.
\end{equation}
However, we can go one step further. For a massive string, corresponding to the
string center of mass fulfilling (in units where $\alpha'=1$):
\begin{equation}
g_{\mu\nu}(q)\dot{q}^\mu\dot{q}^\nu=-m^2,
\end{equation}
there are $D-1$ physical polarizations
of string perturbations around the geodesic $q^\mu(\tau)$. We therefore
introduce $D-1$ normal vectors $n^\mu_R,\;R=1,2,...,(D-1)$:
\begin{equation}
g_{\mu\nu}n^\mu_R\dot{q}^\nu=0,\;\;\;g_{\mu\nu}n^\mu_R n^\nu_S=\delta_{RS}
\end{equation}
and consider only first order perturbations in the form:
\begin{equation}
\eta^\mu=\delta x^R n^\mu_R,
\end{equation}
where $\delta x^R$ are the comoving perturbations, i.e. the perturbations
as seen by an observer travelling with the center of mass of the string. The
normal vectors are not uniquely defined by Eqs. (6.8). In fact, there
is a gauge invariance originating from the freedom to make local
rotations of the $(D-1)$-bein spanned by the normal vectors. For our purposes
it is convenient to fix the gauge taking
the normal vectors to be covariantly constant:
\begin{equation}
\dot{q}^\mu\nabla_\mu n^\nu_R=0.
\end{equation}
This is achieved by choosing the basis $(q^\mu, n^\mu_R)$ obeying the
conditions
given by Eqs. (6.8) at a given point, and defining it
along the geodesic by means of
parallel transport. Another useful formula is the completeness relation
that takes the form:
\begin{equation}
g^{\mu\nu}=-\frac{1}{m^2}\dot{q}^\mu\dot{q}^\nu+n^{\mu R}n^\nu_R.
\end{equation}
Using Eqs. (6.7)-(6.10) in Eq. (6.6), we find after multiplication by
$g_{\mu\nu}n^\nu_S$ the spacetime invariant formula \cite{all1}:
\begin{equation}
(\partial^2_\tau-\partial^2_\sigma)\delta x_R-R_{\mu\rho\sigma\nu}
n^\mu_R n^\nu_S
\dot{q}^\rho\dot{q}^\sigma\delta x^S=0.
\end{equation}
Since the last term depends on $\sigma$ only through $\delta x^S$ it is
convenient to make a Fourier expansion:
\begin{equation}
\delta x_R(\tau,\sigma)=\sum_n C_{nR}(\tau)e^{-in\sigma}
\end{equation}
Then Eq. (6.12) finally reduces to:
\begin{equation}
\ddot{C}_{nR}+(n^2\delta_{RS}-R_{\mu\rho\sigma\nu}n^\mu_R n^\nu_S
\dot{q}^\rho\dot{q}^\sigma)C^S_n=0,
\end{equation}
which constitutes a matrix Schr\"{o}dinger equation with $\tau$ playing the
role of the spatial coordinate. Notice that in the case of constant
curvature spacetimes, $R_{\mu\rho\sigma\nu}\propto (g_{\mu\sigma}g_{\rho\nu}-
g_{\mu\nu}g_{\rho\sigma}),$ the "potential" in Eq. (6.14) is obtained directly
from the normalization equations (6.7) and (6.8) without any calculations
at all.
\vskip 6pt
\hspace*{-6mm}For the
second order perturbations the picture is a little more complicated.
Since they couple to the first order perturbations we consider the full set
of perturbations $\xi^\mu$ \cite{ve3,men}:
\begin{equation}
\dot{q}^\lambda\nabla_\lambda(\dot{q}^\delta\nabla_\delta\xi^\mu)-
R^\mu_{\epsilon\delta\lambda}\dot{q}^\epsilon\dot{q}^\delta\xi^\lambda-
\xi''^\mu=U^\mu,
\end{equation}
where the source $U^\mu$ is bilinear in the first order perturbations, and
explicitly given by:
\begin{equation}
U^\mu=-\Gamma^\mu_{\rho\sigma}(\dot{\eta}^\rho\dot{\eta}^\sigma-
\eta'^\rho\eta'^\sigma)-2\Gamma^\mu_{\rho\sigma,\lambda}\dot{q}^\rho
\eta^\lambda\dot{\eta}^\sigma-\frac{1}{2}\Gamma^\mu_{\rho\sigma,
\lambda\delta}\dot{q}^\rho\dot{q}^\sigma\eta^\lambda\eta^\delta.
\end{equation}
After solving Eqs. (6.14)-(6.15) for the first and second order
perturbations, the constraints, Eq. (6.2), have to be imposed. In world-sheet
light cone coordinates $(\sigma^\pm=\tau\pm\sigma)$
the constraints take the form:
\begin{equation}
T_{\pm\pm}=g_{\mu\nu}\partial_\pm x^\mu\partial_\pm x^\nu=0,
\end{equation}
where $\partial_\pm=\frac{1}{2}(\partial_\tau\pm\partial_\sigma)$.
The world-sheet energy-momentum tensor $T_{\pm\pm}$ is conserved, as can be
easily verified using Eq. (6.1), and therefore can be written:
\begin{equation}
T_{--}=\frac{1}{2\pi}\sum_n\tilde{L}_n e^{-in(\sigma-\tau)},\;\;\;T_{++}=
\frac{1}{2\pi}\sum_n L_n e^{-in(\sigma+\tau)}.
\end{equation}
At the classical level under consideration here, the constraints are
then simply:
\begin{equation}
L_n=\tilde{L}_n=0.
\end{equation}
Up to second order in the expansion around the string center of mass we find:
\begin{eqnarray}
T_{\pm\pm}\hspace*{-2mm}&=&\hspace*{-2mm}-\frac{1}{4}m^2+g_{\mu\nu}
\dot{q}^\mu\partial_\pm\eta^\nu
+\frac{1}{4}g_{\mu\nu,\rho}\dot{q}^\mu
\dot{q}^\nu\eta^\rho\nonumber\\
\hspace*{-2mm}&+&\hspace*{-2mm}g_{\mu\nu}\dot{q}^\mu\partial_\pm\xi^\nu
+g_{\mu\nu}\partial_\pm\eta^\mu\partial_\pm
\eta^\nu+g_{\mu\nu,\rho}\dot{q}^\mu\eta^\rho\partial_\pm\eta^\nu\nonumber\\
\hspace*{-2mm}&+&\hspace*{-2mm}\frac{1}{4}g_{\mu\nu,\rho}\dot{q}^\mu
\dot{q}^\nu\xi^\rho+\frac{1}{8}g_{\mu\nu,\rho\sigma}\dot{q}^\mu
\dot{q}^\nu\eta^\rho\eta^\sigma
\end{eqnarray}
and the conditions $L_0=\tilde{L}_0=0,$ Eq. (6.19), then give a formula for
the string mass.

In the case of ordinary $D\geq 4$ de Sitter space, the first order
perturbations, Eq. (6.14), are determined by \cite{ve3}:
\begin{equation}
\ddot{C}_{nR}+(n^2-m^2H^2)C_{nR}=0.
\end{equation}
For $mH>\mid n\mid$ the frequencies become imaginary and classical
instabilities develop. After solving the second order perturbation
equations, Eqs. (6.15)-(6.16), the mass formula is found \cite{ve3}:
\begin{equation}
m^2=2\sum_n (2n^2-m^2H^2)\sum_R A_{nR}\tilde{A}_{-nR}
\end{equation}
After quantization one finds that real mass states can only be defined up
to some maximal mass \cite{ve3}. This is reminiscent of the classical
string instabilities in de Sitter space. In ordinary anti de Sitter space
and in the $2+1$ black hole Ads (which is locally, but not globally, Ads) the
first order perturbation equations and mass formula take the form of
Eqs. (6.21)-(6.22), but with $H^2$ replaced by $-H^2$ (or $-1/l^2$ depending
on notation). It follows that in these cases there are no instabilities
neither classically nor quantum mechanically. The perturbation series
approach is perfectly well-defined in these cases.

It is interesting to compare the $2+1$ black hole Ads with ordinary $D\geq 4$
black hole Ads. In Secs. 4,5 we found that the circular string motion is very
similar in these backgrounds (see Table II). This was actually somewhat
surprising since the backgrounds are really very different: the ordinary
black holes have a strong curvature singularity at $r=0,$ the $2+1$ black
hole Ads has not. The circular strings are however very special
configurations and that could be the reason why we did not really see any
qualitative differences. For generic strings we would expect some
differences, however, and that is indeed what we find when using the string
perturbation series approach. In the ordinary black hole Ads spacetime the
first order perturbation equations become \cite{all1}:
\begin{equation}
\ddot{C}_{n\perp}+(n^2+m^2H^2+\frac{Mm^2}{r^3})C_{n\perp}=0,
\end{equation}
\begin{equation}
\ddot{C}_{n\parallel}+(n^2+m^2H^2-\frac{2Mm^2}{r^3})C_{n\parallel}=0,
\end{equation}
for the transverse and longitudinal perturbations, respectively. For the
transverse perturbations, Eq. (6.23), the bracket is always positive, thus
the frequencies are real and no instabilities arise, not even for
$r\rightarrow 0.$ For the longitudinal perturbations, on the other hand, the
bracket can be negative. In that case imaginary frequencies develop. The
($\mid n\mid=1$)-instability sets in at:
\begin{equation}
r_{\mbox{inst.}}=\left(\frac{2Mm^2}{1+m^2H^2}\right)^{1/3}
\end{equation}
The higher modes develop instabilities for smaller $r,$ i.e. closer to the
singularity. Similar results are obtained in the black string background
\cite{all1}, see Table III.
\section{Conclusion}
We have studied the string propagation in de Sitter and black hole backgrounds.
The main part of the talk concerned the dynamics of circular strings in de
Sitter space, with special interest in the physical interpretation of the
results of the mathematical analysis. We then compared with results obtained
in various black hole backgrounds ($2+1$ BH-ADS, Schwarzschild-anti de Sitter,
Schwarzschild-de Sitter). Finally, more insight about the strings in curved
spacetimes was obtained using the string perturbation series approach. The main
results and conclusions are summarized in tables I,II,III.
\vskip 24pt
\hspace*{-6mm}{\bf Acknowledgements} \\
I would
like to thank H.J. de Vega and N. S\'{a}nchez
for a very fruitful collaboration on the material presented in this talk.


\newpage
\begin{thebibliography}{11}
\bibitem{ve1}H.J. de Vega and N. S\'{a}nchez, {\it Phys. Rev.} {\bf D42} (1990)
             3969.\\
             H.J. de Vega, M.R. Medrano and N. S\'{a}nchez,
             {\it Nucl. Phys.} {\bf B374} (1992) 405.
\bibitem{ve2}H.J. de Vega and N. S\'{a}nchez, {\it Phys. Lett.} {\bf B244}
             (1990) 215, {\it Phys. Rev. Lett.} {\bf 65C} (1990) 1517,
             {\it IJMP} {\bf A7} (1992) 3043, {\it Nucl. Phys.} {\bf B317}
             (1989) 706 and ibid 731.\\
             D. Amati and C. Klimcik, {\it Phys. Lett.} {\bf B210} (1988) 92.\\
             M. Costa and H.J. de Vega, {\it Ann. Phys.} {\bf 211} (1991) 223
             and ibid 235.\\
             C. Loust\'{o} and N. S\'{a}nchez, {\it Phys. Rev.} {\bf D46}
             (1992) 4520.
\bibitem{mic1}V.E. Zakharov and A.V. Mikhailov, {\it JETP} {\bf 47}
             (1979) 1017.\\
             H. Eichenherr, in {\it Integrable Quantum Field Theories}, ed.
             J. Hietarinta and C. Montonen (Springer, Berlin, 1982).
\bibitem{bar}I. Bars and K. Sfetsos, {\it Mod. Phys. Lett.} {\bf A7}
             (1992) 1091.\\
             H.J. de Vega, J.R. Mittelbrunn, M.R. Medrano and N. S\'{a}nchez,
             "The Two-Dimensional Stringy Black Hole: a new Approach and a
             Pathology", {\it PAR-LPTHE} {\bf 93/14} and "The General Solution
             of the 2-D Sigma Model Stringy Black Hole and the Complex
             Sine-Gordon Equation", {\it PAR-LPTHE} {\bf 93/53}.
\bibitem{ve3}H.J. de Vega and N. S\'{a}nchez, {\it Phys. Lett.} {\bf B197}
             (1987) 320.
\bibitem{men}P.F. Mende, in {\it String Quantum Gravity and the Physics at
             the Planck Scale}, ed. N. S\'{a}nchez (World Scientific,
             Singapore, 1993).
\bibitem{fro1}A.L. Larsen and V.P. Frolov, {\it Nucl. Phys.} {\bf B414}
             (1994) 129.
\bibitem{all1}A.L. Larsen and N. S\'{a}nchez, "Strings Propagating in the
              2+1 Dimensional Black Hole Anti de Sitter Spacetime",
             {\it Obs. de Paris, DEMIRM} {\bf 94013}.
\bibitem{ve5}The contributions by H.J. de Vega and N. S\'{a}nchez in
             {\it String Quantum Gravity and the Physics at the Planck
             Scale}, ed. N. S\'{a}nchez (World Scientific, Singapore, 1993).
\bibitem{ven}N. S\'{a}nchez and G. Veneziano, {\it Nucl. Phys.} {\bf B333}
             (1990) 253. \\
             M. Gasperini, N. S\'{a}nchez and G. Veneziano,
             {\it Nucl. Phys.} {\bf B364} (1991) 365.
\bibitem{fro2}V.P. Frolov, V.D. Skarzhinsky, A.I. Zelnikov and O. Heinrich,
             {\it Phys. Lett.} {\bf B224} (1989) 255.
\bibitem{mic2}H.J. de Vega, A.V. Mikhailov and N. S\'{a}nchez,
             {\it Teor. Mat. Fiz.} {\bf 94} (1993) 232.
\bibitem{all2}H.J. de Vega, A.L. Larsen and N. S\'{a}nchez, "Infinitely many
              Strings in de Sitter Spacetime: Expanding and Oscillating
              Elliptic Function Solutions", {\it Obs. de Paris, DEMIRM}
              {\bf 93055}.
\bibitem{all3}A.L. Larsen, "Circular String-Instabilities in Curved Spacetime",
             {\it Obs. de Paris, DEMIRM} {\bf 93052}, to appear in Phys. Rev.
D.
\bibitem{verb}A. Davidson, N.K. Nielsen and Y. Verbin, {\it Nucl. Phys.}
             {\bf B412} (1994) 391.
\bibitem{all4}H.J. de Vega, A.L. Larsen and N. S\'{a}nchez, "Semi-Classical
              Quantization of Circular Strings in de Sitter and anti de
              Sitter Spacetime", in preparation.
\bibitem{ban}M. Ba\~{n}ados, C. Teitelboim and J. Zanelli,
             {\it Phys. Rev. Lett.} {\bf 69} (1992) 1849.
\bibitem{gra}I.S. Gradshteyn and I.M. Ryznik, {\it Table of Integrals,
             Series and Products} (Academic Press Inc, London, 1980).
\bibitem{abr}M. Abramowitz and I.A. Stegun, {\it Handbook of Mathematical
             Functions} (Dover Publications Inc, New York, 1970).
\bibitem{all5}A.L. Larsen, "Stable and Unstable Circular Strings in
              Inflationary Universes", {\it NORDITA} {\bf 94/14 P}.
\bibitem{ban2}M. Ba\~{n}ados, M. Henneaux, C. Teitelboim and J. Zanelli,
              {\it Phys. Rev.} {\bf D48} (1993) 1506.
\bibitem{wel}G.T. Horowitz and D.L. Welch, {\it Phys. Rev. Lett.}
             {\bf 71} (1993) 328.
\bibitem{ini}H.J. de Vega and I.L. Egusquiza, {\it Phys. Rev.} {\bf D49}
             (1994) 763.
\end{thebibliography}
\end{document}